%% file: MANUSCRIPT.tex
\documentclass[lettersize,journal]{IEEEtran}
\usepackage{amsmath,amsfonts}

\usepackage{algpseudocode}

\usepackage{algorithm}
\usepackage{array}
\usepackage[caption=false,font=normalsize,labelfont=sf,textfont=sf]{subfig}
\usepackage{textcomp}
\usepackage{stfloats}
\usepackage{url}
\usepackage{verbatim}
\usepackage{graphicx}

\usepackage{multirow} 

\usepackage{booktabs} 
\usepackage{tabularx} 
\usepackage{adjustbox} 
\usepackage{lipsum} 

\usepackage[style=numeric, sorting = none]{biblatex}
\begin{document}

\title{Platform-Driven Collaboration Patterns: Structural Evolution Over Time and Scale}

\author{Negin Maddah, Babak Heydari

Department of Mechanical and Industrial Engineering, Northeastern University,

Network Science Institute, Boston, MA 02115, United States

\thanks{Manuscript submitted Feb, 2024}}

\markboth{ieee transactions on computational social systems, Feb~2024}%
{Shell \MakeLowercase{\textit{et al.}}: A Sample Article Using IEEEtran.cls for IEEE Journals}

\maketitle

\begin{abstract}
Within an increasingly digitalized organizational landscape, this research delves into the dynamics of decentralized collaboration, contrasting it with traditional collaboration models. An effective capturing of high-level collaborations (beyond direct massages) is introduced as the network construction methodology including both temporal and content dimensions of user collaborations – an Alternating Timed Interaction (ATI) metric as the first aspect, and a quantitative strategy of thematic similarity as the second aspect. This study validates three hypotheses that collectively underscore the complexities of digital team dynamics within sociotechnical systems: Firstly, it establishes the significant influence of problem context on team structures in work environments, emphasizing the need to consider the specific nature of tasks in analyzing collaborative dynamics. Secondly, the study reveals specific evolving patterns of team structures on digital platforms concerning team size and artifact maturity. Lastly, it identifies substantial differences in team structure patterns between digital platforms and traditional organizational settings, underscoring the unexplored nature of digital collaboration dynamics. The findings of this study are instrumental for organizations navigating the digital era, offering insights into effective knowledge sharing in the decentralized leadership of digital teams. By mapping out network structures and collaborative patterns, this study, with a focus on Wikipedia as a representative digital platform, paves the way for strategic interventions to optimize digital team dynamics and align them with broader organizational goals. 
\end{abstract}


\section{Introduction}

In recent years, the landscape of knowledge sharing (KS) within teams has undergone a significant transformation, driven by advancements in web and digital technologies \cite{sensuse2021exploring, barley2017changing}. Digital platforms, functioning as sociotechnical systems that integrate social communities with technical infrastructure \cite{heydari2018guiding,heydari2020analysis}, have transcended traditional geographical limits, becoming pivotal to innovation and design processes \cite{gilson2015virtual, lee2017pmo, gunasekera2018knowledge, choi2019mechanism, gao2016factors}. These platforms are not merely tools but the core of shaping, mediating, and governing the processes of collaborative design. This shift towards leveraging diverse perspectives and skills, coupled with the emphasis on rapid prototyping within competitive markets, marks a departure from centralized expertise to a more distributed, global collaboration model. Despite recognizing the critical role of such collaborative efforts in ensuring effective governance, particularly highlighted during responses to crises like the COVID-19 pandemic \cite{scognamiglio2023public}, the dynamics of decentralized KS demand further exploration.

Knowledge is a cornerstone of organizational success and economic value creation \cite{sensuse2021exploring, natarajan2008knowledge}, thriving through effective individual interactions within organizations \cite{rahayuab2020preliminary}. The importance of KS for enhancing innovation and work efficiency is well-documented, playing a crucial role across both corporate and educational sociotechnical systems \cite{suto2015representation, page2007making, cox1991managing, paulus2003group, katzenbach2015wisdom}. Co-creation teams are seen as dynamic ecosystems gathered by the common objectives of knowledge, product, or service creation. These teams enable multidimensional interactions that drive innovation by merging diverse knowledge bases \cite{ramaswamy2018co, blomqvist2006collaboration}. Such teams can be studied to investigate the evolving nature of KS dynamics in adapting to new technological landscapes.

The increasing prevalence of online KS activities amplified by social networking platforms \cite{das2018collaboration}, and coupled with the transformative impact of modern technology on corporate culture \cite{babu2008knowledge}, illustrates the critical interplay between the advancement of knowledge management practices and technology adoption. This synergy is vital for sustaining the effectiveness of organizations \cite{gibson2022sustaining}. The distinction between offline and online environments for KS is crucial in understanding the dynamics of traditional versus digital collaboration systems \cite{charband2016online}. This paradigm shift, marked by an increasing reliance on digital decentralized collaboration, is reshaping team interactions, and requires organizations to reevaluate virtual team dynamics, and leadership roles to adapt their strategies and policies to optimize the benefits of digital KS. 

In virtual teams, leaders play a pivotal role in fostering effective communication and managing team challenges \cite{gibson2006unpacking, gilson2015virtual, marlow2017communication}. Peer production teams, exemplified by platforms like Wikipedia \cite{larson2020leading, zhu2012effectiveness, zhu2013effectiveness, kittur2008harnessing, kittur2010beyond}, demonstrate a trend towards emergent, shared leadership, often without formal structure or compensation. This trend has been widely adopted \cite{pena2023categorization, wang2019decentralized, chesbrough2007open} (or aimed to be adopted) in formal organizations for open innovation projects for years, which leads to a transformed leadership in team environments to more decentralized models. Larson and DeChurch highlight how these changes necessitate a reevaluation of leadership dynamics in digital platforms in a decent review study \cite{larson2020leading}. The shift towards decentralized, collective leadership is supported by a body of research \cite{mcgrath1962leadership, zaccaro2001team, morgeson2010leadership, d2016meta, nicolaides2014shared, wang2014meta,heydari2015efficient, mosleh2016distributed, mosleh2016distributed2}, indicating more effective leadership in digital contexts, where traditional hierarchies are less applicable and the need for agile, responsive leadership is more pronounced.
The formation and evolution of teams in digital environments significantly influence leadership dynamics \cite{zhu2011identifying}. Larson and DeChurch highlight that in the context of digital collaborations like Wikipedia, team members' interactions during team formation can be instrumental in determining who emerges as a leader. Wikipedia, the peer-produced online encyclopedia, initially led by its founder Jimmy Wales \cite{wales2001wikimedia}, now operates with a more decentralized leadership model; current Wikipedia page editors play a critical role in decision-making, shaping the direction \cite{larson2020leading} and flow of information on the platform. 

Understanding these transformed dynamics is essential for organizations operating in digital spaces, where traditional leadership models may be less appropriate. Following our digital platform example, Wikipedia, and considering the significant influence that editors wield in shaping its content and structure \cite{larson2020leading}, it becomes pertinent to explore the extensive research focusing on decentralized collaborations. Numerous studies have concentrated on the collaborative dynamics of Wikipedia editors. For instance, \cite{korfiatis2006evaluating} and \cite{brandes2009network} delve into the social network models within Wikipedia, exploring how they influence editorial authority and collaboration. Similarly, \cite{laniado2011wikipedians, maniu2011building, yasseri2012dynamics} provide insights into interaction patterns, signed networks of editor attitudes, and the dynamics of conflicts in Wikipedia. It should be noted that scholarly attention on Wikipedia extends beyond editor collaborations. For example, research on knowledge growth, such as the studies by \cite{lydon2021hunters, zhu2020content}, investigates how information is sought, processed, and expanded within the platform. Additionally, the evaluation of content quality is a critical area of research, with significant contributions from \cite{bassani2019automatically} and \cite{wang2020assessing}, who apply machine learning models to assess the quality of Wikipedia articles. Furthermore, comprehensive reviews of Wikipedia research, such as the recent work by \cite{ren2023did}, provide a valuable overview of the platform's multifaceted nature, encompassing editor behaviors, collaboration processes, and content development. The focus of this research, however, remains centered on the lens of Wikipedia editors as collaborators in the co-creation process to contribute to the broader understanding of collaborations in digital knowledge platforms.

In today’s digital age, understanding the dynamics of information flow is crucial for organizational success \cite{palocs2012information, maddah2023data}. Employing network analysis, this study delves into how team dynamics influence aspects such as information leadership, diffusion of new ideas, cooperation behaviors, and the formation of sub-communities, grounded in network science principles \cite{watts1998collective, burt2004structural, uzzi2005collaboration, singh2010world, gianetto2015network}. By evaluating the network structure across different topics and stages of artifact development, we aim to uncover insights into digital collaboration's complexities \cite{yang2021cooperation, park2020understanding, lazer2007network, mason2008propagation}.

This study proposes a novel framework to analyze digital co-creation, focusing on the high-level, indirect interactions that previous models overlook. Unlike approaches that concentrate on explicit direct exchanges \cite{jacobs2021large, kossinets2009origins}, our method captures the essence of collaboration through subtle, interactive sequential interactions, building upon others’ work over time, that form the backbone of virtual co-creation networks. By introducing an Alternating Timed Interaction (ATI) metric, we measure the immediacy of collaboration within specific time windows, highlighting the importance of temporal proximity. Additionally, our methodology includes a content processing strategy to assess interaction depth by analyzing thematic overlaps in contributors' revisions, offering a comprehensive view of digital co-creation dynamics for different categories of artifacts. This methodology not only illuminates the complex interaction patterns within Wikipedia's editorial network but also serves as a model extendable to other digital co-creation contexts, paving the way for future research and practical applications by enhancing our understanding of digital collaboration's multifaceted nature.

Despite the evident shift towards digital platforms in sociotechnical systems, there remains a significant research gap in fully understanding the complexities of team dynamics in digital collaborative environments. This study aims to bridge this gap by conducting a comparative analysis of “Pure Digital Decentralized Platform-Driven Collaboration” and “Traditional/Hierarchical Organizational Collaboration”. This comparison aims to dissect the interaction networks—whether formal or informal—within these contrasting collaboration models. The post-COVID era of online/hybrid communication provided the opportunity to examine this comparative study. 

Using Wikipedia as a case study, we explore the intricacies of decentralized platform-driven collaboration, analyzing its co-creation networks to reveal the evolutionary dynamics of team collaboration \cite{chen2014community}. Wikipedia facilitates this study with its rich data on diverse topics of articles as artifacts, with different maturity levels, and various collaboration team scales.

To encapsulate our exploration focus, we anchor the study around three critical research questions designed to deepen our understanding of digital collaborative dynamics: 

\begin{enumerate}
    \item {Do collaborators behave differently when co-creating in different contexts of artifacts?}
    \item {How are the dynamics of team structures (such as decentralized leadership, information accessibility, diffusion speed of ideas, and the formation of sub-communities and polarization) evolving over scale and time?}
    \item {How different are the structural patterns of teams within a pure digital collaboration environment from those in traditional organizational collaborations?}
\end{enumerate}

This research is poised to delve into the intricate dynamics and challenges of digital co-creation, seeking to contribute valuable insights into its unique characteristics. The subsequent sections of this paper are organized as follows: we begin with a detailed explanation of our methodology, followed by its application to the study of editorial collaborations on Wikipedia. We then present an analysis of our findings and address the research questions posed earlier. The paper concludes with a discussion of our study, its limitations, and directions for future research.

\section{Methodology}
In line with our exploration of digital collaboration dynamics for a comparative analysis of them with traditional organizational structures, we draw inspiration from recent studies that have performed extensive data analysis of organizational collaboration in hybrid and online environments. One such study meticulously done by Jacobs and Watts involves an exploratory analysis of a unique dataset comprising 1.8 billion messages sent by 1.4 million users from 65 publicly traded U.S. firms across various sizes and industrial sectors, investigating collaboration dynamics in contemporary settings \cite{jacobs2021large}.
This precedent of analyzing large-scale, real-world data sets in organizational contexts provides a robust benchmark for our investigation. Our approach initially focuses on conducting a detailed network analysis of a purely digital platform. By examining the interactions to co-create within this digital environment, we aim to derive insights that can be systematically compared with the patterns observed in traditional organizational collaborations.

\subsection{Domain Selection of the Artifact}

In aligning our methodology with the multifaceted nature of digital co-creation platforms, this study strategically selects data across five broad categories, reflecting the insights of Gibbs, Sivunen, and Boyraz \cite{gibbs2017investigating} regarding the significant influence of task nature in virtual teams. These categories include (1) Politics, capturing discussions and collaborations surrounding key individuals in the political arena such as 'Ron DeSantis' and 'Alexandria Ocasio-Cortez' and political events such as 'Impeachment inquiry against Donald Trump'; (2) International Conflicts such as 'Hong Kong protests' and 'Russo-Ukrainian War’ encompassing topics related to global conflicts to reflect the complexity of collaborative content in areas of international tension; (3) Natural and Man-Made Disasters, which includes a range of disaster-related topics highlighting collaborative efforts during global crises like 'COVID-19', '2018 California wildfires', and '2020 Beirut explosion'; (4) Technology and Business with articles such as 'Tesla Model 3' and 'Cryptocurrency', focusing on the rapidly evolving sectors that significantly influence digital collaboration patterns; and finally, (5) Entertainment including  'Parasite (2019 film)' and ‘Kylian Mbappé’, covering a spectrum of topics to demonstrate the diversity of interest and engagement in collaborative platforms. A full list of the sample articles used in this paper exists in the first section of the Appendix Document.

While this selection is tailored to our study's context using Wikipedia, it should be adapted and specialized based on the application and nature of tasks in other digital platforms within each organization. This approach ensures a balanced analysis suitable for broader applicability and adaptability in future digital co-creation explorations.

{\bf{Hypothesis 1: Collaborators behave differently when co-creating in different contexts of artifacts.}}. {\bf{In other words, the nature of the problem impacts the network formation and evolution in digital platforms. }}. In proposing this hypothesis, we aim to investigate whether the specific context of an artifact—be it politics, international conflicts, disasters, technology, or entertainment—has a discernible impact on the structuring and evolution of collaborative networks. This investigation acknowledges the potential for significant variability in how collaborative efforts manifest across different contexts within organizations.

Additionally, we deliberately select articles predominantly post-2010. We avoided older articles to minimize bias that could arise from Wikipedia's early developmental stages. This ensures that our analysis is reflective of more mature and established collaborative patterns on the platform. Figure 1 is a visual representation of our domain selection, showcasing the distribution of the sample articles across each category and highlighting the creation dates of these articles. 

\begin{figure}[!t]
\centering
\includegraphics[width=3.4in]{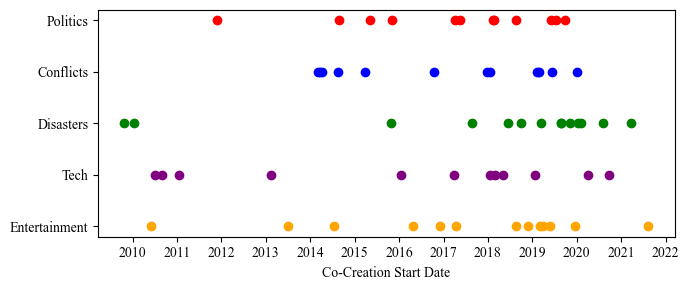}
\caption{The temporal distribution and the number of sample topics selected for each category. This figure facilitates an understanding of the temporal spread and categorical diversity of the sample artifacts studied.}
\label{fig_1}
\end{figure}

\subsection{Data Acquisition}

A detailed investigation of team dynamics and information flow in digital co-creation environments is facilitated by the online traceability of collaboration steps, contrasting with the opaque nature of in-person teams, where most of the details of collaboration may remain unrecorded. In organizational contexts, while data might be anonymous to protect privacy, essential elements such as unique identifiers for each collaborator, time logs of contributions, and content topics of each task are generally accessible. This ensures a comprehensive understanding of how virtual teams operate, evolve, and lead the co-creation process. However, data acquisition, while straightforward in principle, requires meticulous processing. It involves not just the collection but also the careful parsing and interpretation of digital interactions to accurately reflect the dynamics of collaboration and co-creation. 

The data acquisition phase for our study involved collecting revision histories from selected Wikipedia articles. This process was facilitated through the Wikipedia API; we captured each editor's unique identifier to map individual contributions and participation patterns within the editorial network. Additionally, timestamps of each edit were recorded to analyze the temporal dynamics of collaboration and observe how interactions evolve. Lastly, detailed information about the content of each revision was collected. 

\subsection{Network Formation of Collaborators}

To decipher the dynamics within a digital platform, our study employs a methodical approach to forming networks among collaborators. For each selected artifact, we establish a specific timeframe, measured in months, to generate a network of interactions. Notably, the commencement of this timeframe is not defined as the artifact's creation date but rather as the onset of co-creation activities; this initiation point varies, ranging from immediate post-creation engagement to several years of latency. Such differentiation ensures that our analysis starts from a period of active co-creation, yielding a more accurate depiction of the collaboration dynamics.
Networks of collaborators for each article within the predefined timeframe are constructed as follows: nodes are represented by the unique IDs of the collaborators, and edges are defined by two principal dimensions - temporal and content interactions among editors:

{\bf{Temporal Dimension:}}. An interaction time threshold (e.g., a few hours) is set as a meta-parameter of our model. Within this parameter, reciprocal interaction activities between two collaborators are quantified. To achieve this, we introduce a measure of mutual interaction, accounting for indirect interactions, which are evidenced by users building upon each other's edits or contributions. Addressing the challenge of inferring indirect interactions, we will use the notion of "Alternating Timed Interactions" (ATI). This pattern, characterized by alternating activities between individuals (e.g., i → j → i → j → i → j) within the specified time window, serves as an indirect indicator of active and reciprocal engagement. It stands in contrast to non-alternating patterns (e.g., i → i → i → j → j → j). Specifically, if a sequence of edits by 'i' and 'j' occurs close enough (within the defined time threshold) in the artifact, this interaction accrues a temporal weight of 1 in the connection between these two editors. This weight increases proportionally with the frequency of such sequential interactions within the defined threshold. 

{\bf{Content Dimension:}}. To quantify the content dimension of interactions among Wikipedia editors, we analyze the thematic similarities between their revisions. This involves identifying the main topics of an article's subsections and employing natural language processing (NLP) techniques to preprocess and analyze the text of revisions for thematic content. By preprocessing the text—tokenizing, removing stopwords, and filtering non-alphabetic tokens—we distill the revisions to their informative essence.
We then match the processed text to article topics using a topic-allocation strategy, assigning scores based on the occurrence of topic-related keywords within the revisions. For each pair of editors i and j working on the same artifact, we determine the overlap in topics they have contributed to by comparing the sets of topics associated with their revisions. The intersection of these sets indicates their thematic alignment. The strength of the connection between two editors is measured by the size of this intersection, normalized by the total number of sub-sections in the article to account for variations in topic breadth. This normalized weight represents the content-based connection strength between editor pairs, offering insights into the thematic coherence of their collaborations.

Given the definitions, the total weight of the connection between collaborators $i$ and $j$ in artifact $a$, denoted by \(W_a^{(i,j)}\), is calculated as the product of the temporal dimension \(W_T^{(i,j)}\) and the content dimension \(W_C^{(i,j)}\) of these weights:

\begin{equation}
W_a^{(i,j)} = W_T^{(i,j)} \cdot W_C^{(i,j)}
\end{equation}

where,

\begin{equation}
W_T^{(i,j)} = \sum_{k=1}^{\left|\tau\right|-1} I\left(u_k \neq u_{k+1}, \left|\tau_{k+1} - \tau_k\right| \leq \bar{t}\right)
\end{equation}

Let \(\tau\) be the set of timestamps of edits for both collaborators and \(\bar{t}\) be the interaction time threshold. Then \(\tau_k\) and \(\tau_{k+1}\) are consecutive timestamps in \(\tau\), and \(u_k\) and \(u_{k+1}\) are the corresponding users for these timestamps. \(I(.)\) is an indicator function that is one if its argument is true and 0 otherwise.

For the content weights, if we denote the set of sections of each artifact \(a\) as \(S_a\), and the subsets of these sections that each collaborator \(i\) has worked on as \(S_i^a\), then the content weight \(W_C^{(i,j)}\) for editors \(i\) and \(j\) in artifact \(a\) is the proportion of overlapping sections they worked on, given by:

\begin{equation}
W_C^{(i,j)} = \frac{|S_i^a \cap S_j^a|}{|S_a|}
\end{equation}

To ensure the analysis focuses on meaningful interactions within the digital platform, the study employs a crucial step of network pruning. This process involves setting a threshold for the minimum weight required for an edge to be considered significant. By filtering out less impactful collaborations, the process distills the network to its most influential connections, offering a clearer view of the core collaborative interactions within each digital artifact. All these steps are summarized in Algorithm 1.

\subsection{Network Structure Analysis}

In the next step, we employ statistical network analysis techniques. While there are plenty of network statistics available, in our approach, we have restricted our statistical network analysis to a selected set of network measures, carefully chosen not only for their ability to capture essential aspects of collaborative behavior but also to allow for a balanced comparison with existing studies on traditional hierarchical organizational structures.

To understand the extent and nature of collaboration, we focused on measuring team cohesion by calculating the average number of interactions each member has within the network. This aspect is crucial as it sheds light on the potential for knowledge transfer and potential innovation \cite{reagans2001networks, reagans2004make, obstfeld2005social, lazer2007network}. Another critical dimension we examined was assessed by determining the average number of steps required to connect any two members of the network, providing insights into the network's capability to facilitate the rapid spread of information \cite{watts1998collective, kogut2001small, uzzi2005collaboration, centola2007complex}. Additionally, we explored the tendency of network members to form closely-knit groups. This measure is indicative of the level of collaboration and open communication within the network, highlighting the potential for collaborative efforts to be concentrated among certain groups \cite{lazer2007network, mason2008propagation, mason2012collaborative, muller2019effect}. Finally, we looked at the network's structural propensity to centralize information flow. By assessing which members predominantly control or influence the flow of information, we gained insights into the network's architecture regarding how it enables certain individuals to act as crucial connectors, impacting the overall communication process \cite{borgatti2003network, monge2003theories, katz2004network, sasidharan2012effects, becker2017network}.

\begin{algorithm}[H]
\caption{Co-Creation Network Construction}
\begin{algorithmic}[1] 

\For{each topic in sample artifacts}
    \State Determine 6-month time windows.
    \For{each time window}
        \State Retrieve edit history for the topic within the
        \State time window.
        \State Initialize all\_weights as an empty dictionary.
        \For{each user pair in edit history}
            \State Combine and timestamp-sort edits to form 
            \State all\_revisions.
            \State Initialize $W_T^{(i,j)}$ and $W_C^{(i,j)}$ to 0 for each pair.
            \For{each pair of adjacent revisions}
                \If{users differ and worked within the \State time threshold}
                    \State Increment $W_T^{(i,j)}$ and calculate $W_C^{(i,j)}$.
                \EndIf
            \EndFor
            \State Store $W_T^{(i,j)}$ and $W_C^{(i,j)}$ in all\_weights.
        \EndFor
    \EndFor
\EndFor

\For{each topic and corresponding time window}
    \State Initialize an empty graph.
    \For{each pair in all\_weights}
        \State Calculate adjusted weight $W_a^{(i,j)}$.
        \If{edge exists}
            \State Update weight.
        \Else
            \State Add new edge with weight.
        \EndIf
    \EndFor
    \State Calculate pruning threshold based on graph metrics
    \State Prune the graph to remove weak connections.
    \State Store the graph for further analysis.
\EndFor

\end{algorithmic}
\end{algorithm}

In line with this analytical framework, we propose our next hypothesis as below.

{\bf{Hypothesis 2: The characteristics of team structures within collaborative networks exhibit discernible patterns that evolve over scale (size) and time (age).}}. This hypothesis aims to explore the dynamic interplay between team structures and the evolving nature of digital collaboration.

Following the insights provided by Jacobs and Watts in their comprehensive 2021 study, "A large-scale comparative study of informal social networks in firms," we find a robust foundation for our analysis. Jacobs and Watts' pioneering work, which involved a detailed analysis of anatomized email data across U.S. firms, revealed notable variations in network metrics influenced by organizational size. Their study highlighted the interplay between micro-level network structures and overarching organizational properties, even though clear correlations with firm age, and the industry that the firms work in were not established.

The strategic alignment of network analysis with existing literature enables us to draw a parallel and conduct a meaningful comparison between the network patterns observed in digital platform-driven collaboration and those prevalent in traditional organizational settings. In this regard, we introduce our final hypothesis as follows:

{\bf{Hypothesis 3: The structural patterns of networks within pure digital collaboration environments, such as Wikipedia, will exhibit significant differences when compared to those found in traditional organizational collaborations.}}. This hypothesis stems from the premise that the inherent characteristics of digital platforms—such as their decentralized nature, the scalability of interactions, and the digital traceability of collaboration processes—cultivate distinctive network structures. These structures are likely not to align with those formed within the more bounded and hierarchical contexts of traditional organizations. By investigating this hypothesis, we aim to shed light on the unique dynamics of digital collaborative networks and how they diverge from conventional organizational network patterns, thereby contributing to a deeper understanding of digital co-creation processes.

In a nutshell, Figure 2 will present an overarching view of the methodology, visually summarizing the steps from domain selection and data acquisition to network formation and calculating network characteristics. This big-picture representation aids in comprehending the comprehensive process undertaken in this study.

\begin{figure}[!t]
\centering
\includegraphics[width=3.3in]{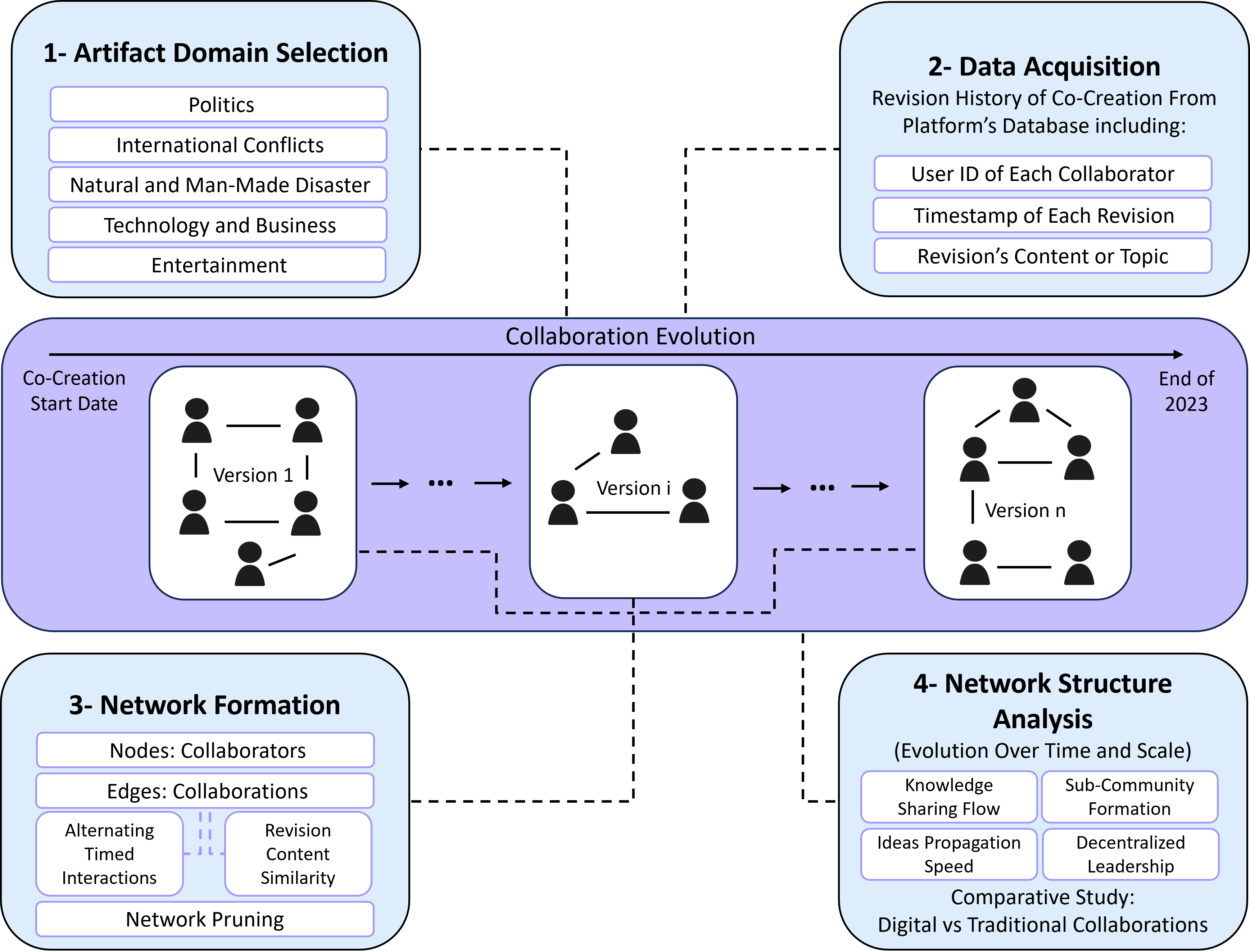}
\caption{Methodology Overview}
\label{fig_2}
\end{figure}

\section{Implementation and Results}

For this study, we constructed interaction networks for Wikipedia collaborators at six-month intervals, starting from the collaboration onset of each article until the end of 2023. Consequently, depending on when each sample artifact was created, we generated multiple collaboration networks for every article, each representing a distinct six-month period. After this, the ATI and content-based weights ($W_T^{(i,j)}$ and $W_C^{(i,j)}$ with the averages of $1.15$ and $0.72$ respectively for the interaction time threshold of $48$ hours) for all the $29,149$ pairs existing in all the sample artifacts are calculated, and the $6$-month period collaboration networks are generated based on them. More detailed information regarding these weights exists in the second section of the Appendix Document. Following the construction of these initial networks, we refined our analysis through a process of network pruning. By setting a threshold for the minimum weight required for an interaction to be considered significant, set at $70\%$ in our analysis, we focused on the most impactful and substantial collaborative interactions. This pruning process of the bottom $30\%$ of the weights was pivotal in ensuring that our networks accurately reflected the core dynamics of collaboration within each digital artifact. Our study further concentrated on analyzing the giant connected component (GCC) of these networks formed, recognizing its critical role in representing the primary structure of collaborative interactions within Wikipedia. 

An additional step of our methodology involved filtering pre-active time windows from the network data: we considered a network as significant when its size exceeded $15\%$ of its maximum size throughout its entire lifetime for that article. There exist some samples for this time-filtering, the reasoning behind choosing this threshold for our dataset, and also the statistical summary of the data before this filtering in the Appendix document. In addition to these steps, we also assessed the maturity of each network by calculating the age of the artifacts in months, from the start of collaboration to the onset of each network formation. 

After filtering the pre-active time windows out, our refined dataset comprised $789$ out of the original $839$ networks, spanning various topics like politics, conflicts, disasters, technology, and entertainment (originally included $168, 155, 145, 192,$ and $179$ respectively). This consolidation resulted in a dataset forming the basis for our subsequent analysis. A comprehensive summary of these networks, including data such as the unique number of topics, network counts, and the range of nodes in the GCCs for each category, is presented in Table ~\ref{tab:table1}. This table provides a snapshot of our dataset's breadth and diversity, offering insights into the scale and temporal spread of the articles analyzed.

\input{table1}

The dataset of collaboration networks also reveals that, on average, they comprise 10 nodes. The average degree (D) stands at 4, indicating a moderate level of interconnectedness among editors With a clustering coefficient (C) of 55\%, the networks tend to form tight-knit groups. The networks exhibit an average of 1.6 for the average shortest path length (L), facilitating efficient information flow, while the average centralization (BC) is less than 1\%, suggesting a decentralized network structure. The age of these networks spans a broad range, from as recent as 0 months to as mature as 158 months. Additionally, it should be noted that the number of edges in the GCC over the number of edges in the original pruned non-connected network, varies between 50\% and 100\%, with an overall average of 97\%. This high ratio underscores the robustness of the GCC in our analysis. For a detailed summary of statistics of the results and a visual presentation of the distribution of these characteristics, please refer to the Appendix Document.

It should also be noted that networks with fewer than 4 nodes are excluded from all our subsequent visualizations and analysis. This decision was based on the rationale that very small networks might not accurately represent the collaborative patterns we sought to study.

\subsection{Network Dynamics and Size of the Team}

As the starting point of our exploration, Figure 3 plots various team structure characteristics against the increasing team size across all categories. This exploration is enriched by the regression model results presented in Table~\ref{tab:size-table}, which examines the relationship between the team size and each of its structure characteristics.

\begin{figure}[!t]
\centering
\includegraphics[width=3.4in]{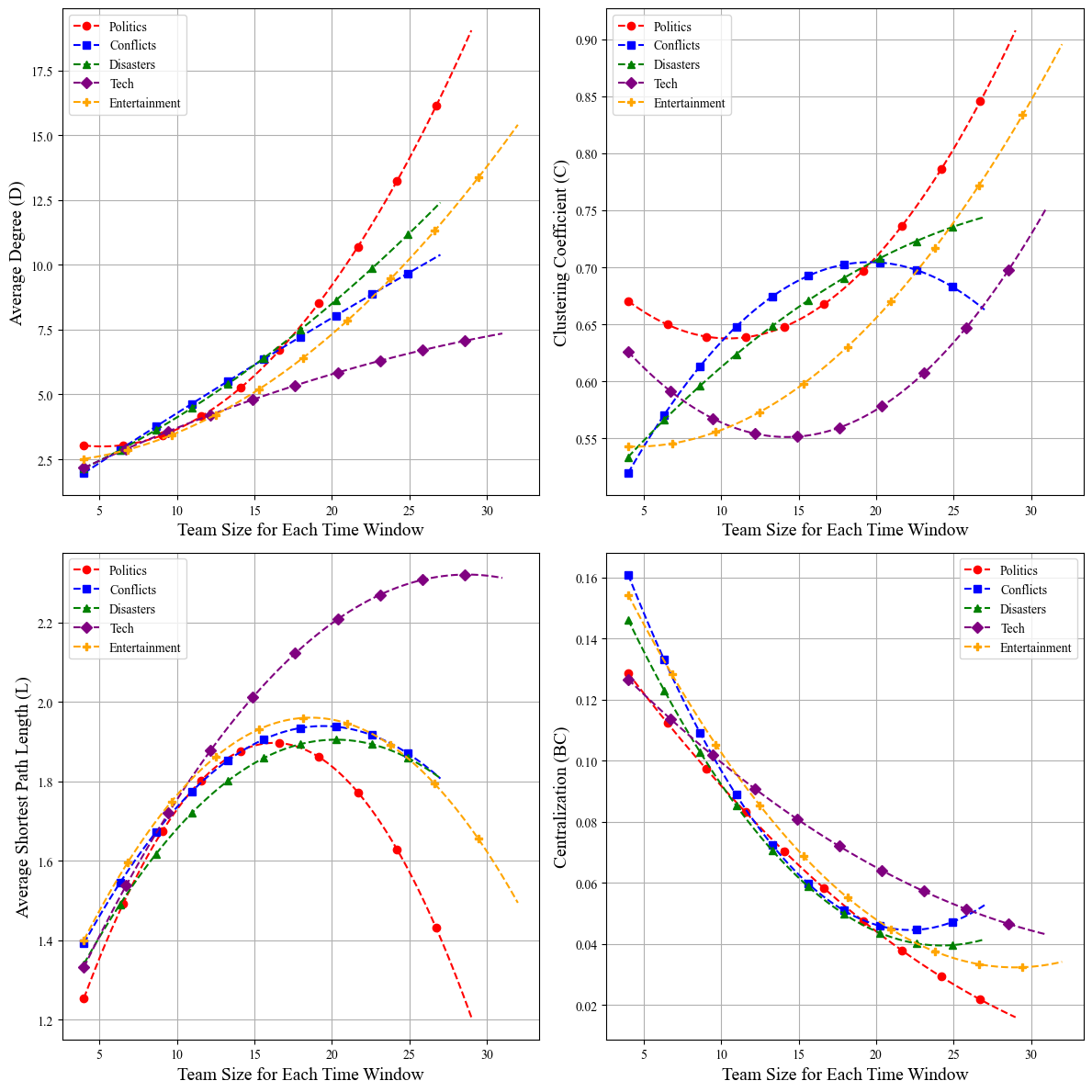}
\caption{Network Structure Patterns When the Teams Grow in Size per Artifacts Category: each collaboration team is defined per 6-month time windows from the onset of the meaningful interactions until the end of 2023.}
\label{fig_3}
\end{figure}

\input{tablesize}

The in-depth regression analysis examining the relationship between network measures and the size of the teams revealed insightful patterns. Notably, in some categories, we observed a lack of significant relationships between some of the network characteristics and the team size, underscoring the influence of content specificity on our results. However, most categories displayed significant associations, highlighting how the nature of each artifact shapes collaborative dynamics. Across all categories and characteristics, our analysis unveiled a spectrum of relationships with team size. Predominantly, these relationships are non-linear, as evidenced by the significance of both linear and squared terms in the regression models. The degree of explanation, represented by R-squared values, varied across characteristics and categories, indicating that some models provided a better fit to the data than others. A visualization of these trends including the scatter plots of the team's network data points exists in the Appendix Document.

{\bf{Team Cohesion and Interconnectivity vs Team Size: }}.

When examining the relationship between interconnectivity captured by the average degree of the team and the number of members collaborating, two perspectives emerge from existing literature: first, Jacobs and Watts' Perspective posits that the average degree does not vary with size which is validated by their massive organizational data analysis; it is based on the assumption that while individual cognitive capacities may vary, the overall average should remain relatively stable across organizations \cite{jacobs2021large}. Secondly, an alternative viewpoint, contrary to Jacobs and Watts, suggests that the average degree should increase with organizational size due to greater availability of potential contacts \cite{leskovec2007graph} or flatter managerial structures \cite{simon1962architecture}.

Our findings align with the second viewpoint for mainly two reasons pertinent to digital platforms: (1) their artifact-centered design: digital platforms are structured to focus on specific artifacts, allowing individuals to concentrate on one problem at a time. This design potentially enhances cognitive capabilities during interactions, as attention remains focused on a singular issue. (2) their decentralized leadership: in contrast to traditional hierarchical organizational structures, digital platforms often feature decentralized leadership. This autonomy allows for parallel interactions with potential collaborators, facilitating an increase in each team member's degree as more people engage with the same problem. To summarize, the connectivity of each team member is given both the opportunity and authority or permission to increase when more people are available to work on the same problem.

This increasing trend when a team grows in size in digital platforms — whether it be a significant quadratic relationship as observed in the political domain, a significant linear relationship as in conflicts and disasters, or more complex relationships in technology and entertainment domains — contrasts with traditional organizational structures. It presents an opportunity for the future of co-creation platforms, where increased interaction links could mean easier information exchange. However, caution is warranted: excessive knowledge sharing might be detrimental, leading to fast convergence on solutions, especially for complex artifacts. This balance between beneficial and excessive information exchange is critical for the effective functioning of collaborative platforms \cite{padhee2023design} (Heydari et al., 2022).

{\bf{Sub-Community Formation and Polarization vs Team Size}}.

In this section, we investigate the intricate relationship between team size and the tendency of sub-community formation, as indicated by clustering patterns, within digital collaborative platforms. This analysis draws upon various network science models, contrasting their predictions with our empirical observations across different topic categories.
Initially, models such as Random Graph Theory \cite{watts1998collective} suggest a decrease in sub-community clustering with increasing team size. In contrast, some Small-World Models \cite{watts1998collective, holme2002growing} and those assuming hierarchical structures \cite{newman2003structure} propose a constancy in clustering despite team growth. However, the scaling proposed by Jacobs and Watts (2021) offers an intermediate view, suggesting that clustering does not fit neatly into a constant or inversely proportional pattern. In our analysis of digital platforms, we observed distinct trends in clustering across different topic categories, diverging from these traditional models:

\begin{itemize}
\item{Conflict and disaster topics: Here, we observed a consistent increase in sub-community clustering, leveling off in very large teams. This trend is likely influenced by the time-sensitive and critical nature of these topics.}
\item{Politics and technology categories: A U-shaped pattern was evident, with initial dispersal in sub-communities followed by a sharp convergence into tightly knit clusters as team size increases.}
\item{Entertainment domain: Although an increasing trend in clustering was observed, it did not present significant patterns, as detailed in Table 3, suggesting a more fluid team structure in this domain.}
\end{itemize}

This observed pattern contradicts traditional network science literature, including random graph models, small-world models, and the findings of Jacobs and Watts. Significantly, the range of clustering coefficients in Wikipedia networks (mostly between 0.5 and 1) is higher than those observed in \cite{jacobs2021large} with a maximum of 0.3, suggesting that digital platforms foster more clustering as teams working on a single artifact within a defined context grow larger. 

The observed trend of increasing clustering in virtual teams, especially as they scale up, likely mirrors the dynamics of sub-community formation and polarization within digital platforms; in larger virtual teams, individuals tend to form clusters around specific interests, expertise, or viewpoints. This is especially pronounced in domains like politics, where the breadth of the subject matter can lead to specialized clusters. The U-shaped pattern observed in politics suggests that as more individuals join the team, they initially explore various aspects of the topic independently before eventually forming tight-knit clusters around specific sub-topics or approaches. On the other hand, in certain contexts, especially in conflicts and disasters, the urgency and high stakes can intensify polarization within teams. Members may cluster around specific solutions or approaches, driven by their experiences, expertise, or emotional responses to the situation. Such polarization, while potentially limiting diverse viewpoints, can also lead to a more in-depth exploration of different strategies.

In a nutshell, as virtual teams grow, the complexity and diversity of the topics lead to more pronounced sub-group formations. This has been one of the most surprising findings of this study. The authors believe that understanding this trend is crucial for organizations leveraging digital platforms for collaboration. While forming clusters can enhance focus and depth in specific areas, it also poses the risk of creating echo chambers that might suppress innovation and diverse perspectives. 

{\bf{Information Propagation Speed vs. Team Size}}.
Previous studies in network science have consistently suggested that the speed of information propagation captured by the average shortest path length (L) increases logarithmically with team size. As Newman (2003) posits, in networks where the average degree (D) is constant, L is expected to rise logarithmically with the number of team members \cite{newman2003structure}. Jacobs and Watts further elaborate that L is highly negatively correlated with D, especially in large firms with hierarchical team structures. In such environments, despite the growth of the firm, if individuals maintain a constant number of contacts, the paths between people stay constant \cite{jacobs2021large}.

However, our analysis reveals a distinct pattern that departs from these conventional findings: for all the categories, we observe a complex relationship; initially, there is an increase in L for mid-size teams, followed by a decrease in L as the network size expands significantly. This pattern aligns with the logic of previous studies. Given that D increases in our data (as elaborated earlier), we anticipated a sharp decline in L towards the end, as the paths between two people should become shorter due to the higher degree of the members and the negative correlation between L and D (around $40\%$ in our dataset). However, it should be noted that in conflicts, disasters, and technology domains, we witness a slower final decrease in L; this observation is intriguing, particularly considering the simultaneous increase in both D and C as team size grows. While several factors could contribute to this trend, one compelling explanation aligns with our earlier hypothesis regarding polarization within these domains.

In summary, the trend we uncover in the speed of information diffusion versus the team size sheds new light on the dynamics of digital collaborative networks, highlighting deviations from established network science theories. The divergence in trends across different categories emphasizes the impact of domain-specific factors on virtual team structures. These insights are crucial for devising effective digital collaboration strategies that are sensitive to the unique characteristics of each domain, ensuring efficient communication and collaboration pathways are maintained regardless of team size or the specific nature of the collaborative endeavor.

{\bf{Centralized Leadership Among Sub-Groups vs Total Team Size}}.
Jacob and Watts' study presents an intriguing insight into organizational network structures, suggesting that centralization remains constant as firms grow, typically the value of 0.85 for betweenness centrality. They also note the ambiguity in prior literature regarding the expected variation of centralization with firm size \cite{jacobs2021large}.
Contrastingly and interestingly, our study unveils a unique significant pattern in the realm of digital collaboration. We observe that the betweenness centrality (BC) in networks formed by digital platform collaborators is not only lower than 0.35 for all the constructed networks during the entire timeline of the study for all the sample artifacts but also tends to decrease as the team size increases. It should be noted that the visual declining trend in BC is not statistically significant in the technology domain but is still very low.

Several factors could potentially explain this difference from the previous studies: (1) Nature of Organizations: Jacob and Watts' research focuses on traditional, hierarchical firms where centralization often pertains to decision-making being concentrated at higher levels. In contrast, Wikipedia's editing environment represents a digital, open-source platform that inherently fosters a distributed collaboration. This fundamental difference in organizational structure could significantly impact centralization metrics. (2) Type of Collaboration: In traditional firms, specific mechanisms may be implemented to sustain a level of centralization as the organization grows, ensuring coherence and strategic direction. Conversely, Wikipedia exemplifies a decentralized model where editors contribute autonomously without a centralized authority overseeing each edit. As such networks expand, the necessity for central nodes diminishes, leading to a decrease in BC. (3) Cultural and Behavioral Factors: The culture of collaboration in open-source communities, such as Wikipedia, emphasizes distributed, egalitarian participation. This cultural inclination towards decentralization contrasts with the more hierarchical and centralized control often seen in corporate environments.

In essence, our findings about the average centrality of leadership between various sub-groups and team size on digital platforms like Wikipedia challenge traditional notions of network centralization, highlighting the distinct dynamics of digital collaborative environments. This divergence not only reflects the structural and cultural differences between digital platforms and traditional organizations but also underscores the importance of considering these unique dynamics when designing and managing digital collaborative systems.

\subsection{Network Dynamics and the Maturity of the Artifact}
In addressing the relationship between team structures and the age or maturity of artifacts in digital collaboration, this study ventures beyond the assertion by Jacobs and Watts that "Network Heterogeneity Is Not Explained by Firm Age" \cite{jacobs2021large}. Aligned with their direction, yet advancing a step further, our investigation pivots to explore how the age of the artifacts, and not the organization, influences the structural dynamics of the teams working on them. This exploration is grounded in the premise that the lifecycle of an artifact—its evolution from inception to maturity—plays a pivotal role in shaping the collaboration patterns and network structures surrounding it.
Our comprehensive analysis across various domains has revealed intricate relationships between the age of artifacts and multiple aspects of team dynamics. Each aspect - team cohesion, sub-community formation, information propagation, and centralized leadership - shows unique patterns of interaction with the artifact's age, reflecting the evolving nature of collaborative efforts. Based on the detailed analysis, it becomes evident that the relationship between team structures and artifact maturity varies across different content categories as shown in Figure 4. This section of the paper delves into the details of these relationships, examining the extent to which the age of an artifact influences the structure of the collaborative networks on Wikipedia. The regression model results of this section are presented in Table ~\ref{tab:age-table}.

\begin{figure}[!t]
\centering
\includegraphics[width=3.4in]{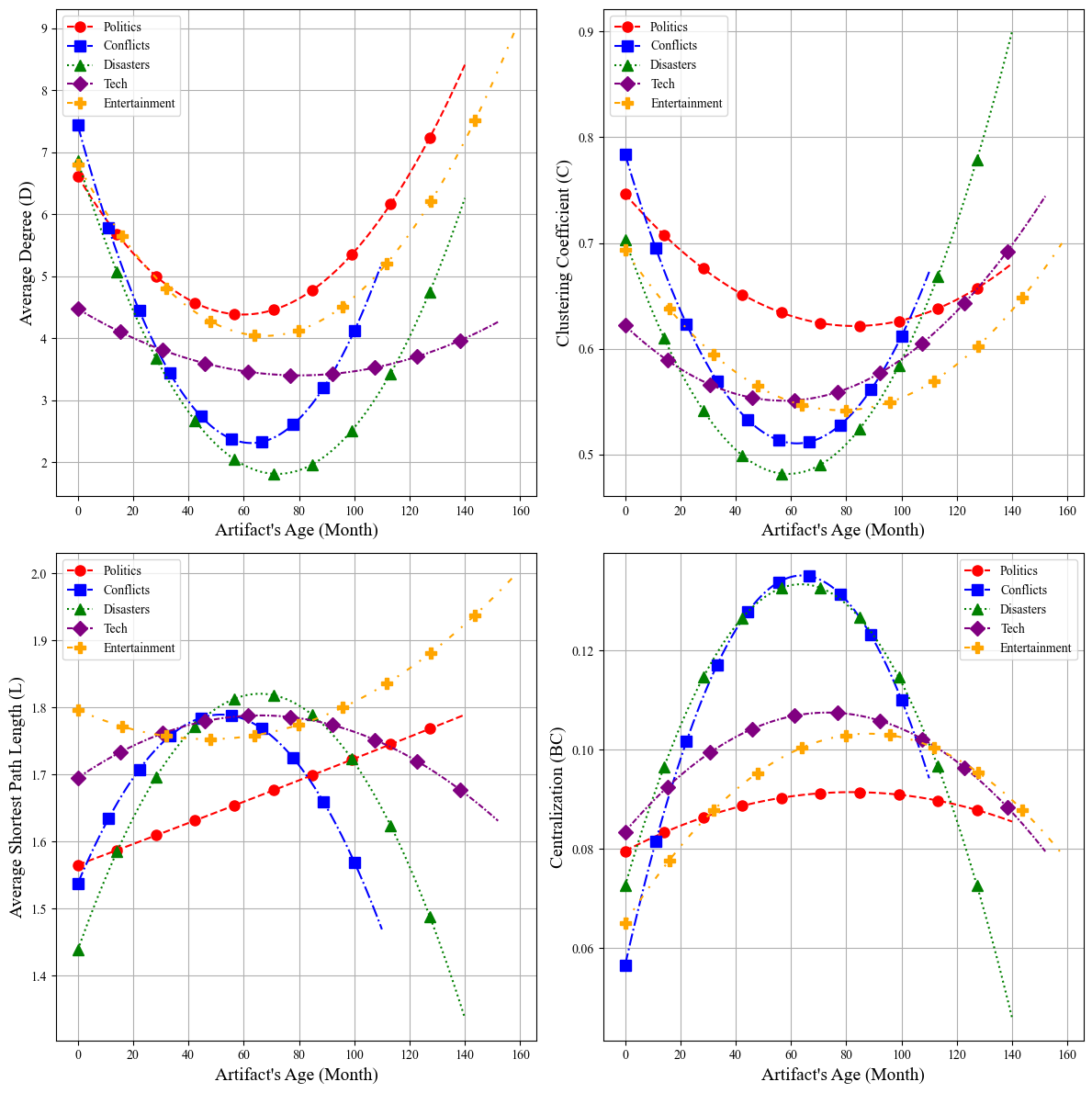}
\caption{Network Structure Patterns When the Artifacts Age For Each Category}
\label{fig_4}
\end{figure}

\input{tableage}

{\bf{Team Cohesion and Interconnectivity vs Artifact’s Age:}}.
In the realm of politics, we observed a subtle yet significant correlation between the depth of collaboration and artifact age. Conversely, in the conflicts and disasters domains, a more pronounced relationship suggests that the maturity of artifacts plays a more substantial role in shaping collaborative dynamics. Across all categories, teams formed around newer articles exhibited higher interconnectivity, which tended to decrease with age, only to increase again in later stages. Notably, the technology domain stood apart, showing no meaningful relationship, likely due to the rapid evolution characteristic of this field.

{\bf{Sub-Community Formation and Polarization vs Artifact’s Age:}}.
The clustering tendency's correlation with artifact age varied across domains. Except for technology, all categories showed a significant, moderately strong correlation. Typically, articles began with high clustering, which decreased over time, and then increased as the artifacts matured. This pattern, which is interestingly similar to the previous pattern, is indicative of the dynamic evolution of information and understanding over time.

{\bf{Information Propagation Speed vs Artifact’s Age:}}.
In terms of information diffusion speed, captured by the average shortest path length, significant correlations with artifact age were only found in the domains of disasters and conflicts, exactly opposite to their average degree and clustering formation patterns. The statistically insignificant relationships among the rest of the domains suggest that while an artifact's maturity might be a factor for some artifacts, it interacts with a multitude of other elements influencing how quickly ideas spread within these teams.

{\bf{Centralized Leadership Among Sub-Groups vs Artifact’s Age:}}.
Finally, the relationship between centralized leadership within sub-groups and the artifact age was significant in the disasters, conflicts, and entertainment categories showing a sharp increase and then a decrease in the centralization of the collaboration teams. However, its explanatory power was somewhat limited, indicating that while artifact age affects network centralization, it is not the only determinant. 

These findings reveal an interesting dynamic between the evolution of collaborative efforts and the maturity of artifacts, particularly pronounced in the conflicts and disasters categories. In these domains, a distinct pattern emerges, characterized by a high level of initial engagement, as indicated by increased team interconnectivity and group cohesiveness, alongside efficient information propagation and centralized decision-making. This heightened activity in the early stages can be linked to the urgent, time-sensitive nature of these topics, fostering immediate and intense collaboration. As events unfold and the immediacy drops, a noticeable shift occurs. We see a decrease in both team interconnectivity and group cohesiveness, along with a lengthening of information propagation paths and a reduction in centralized decision-making. This change reflects a transition in the collaborative focus, moving away from the immediate response to a more prolonged and evolving engagement with the topic. However, as the artifacts grow older, network activity is resurgent. This is marked by a renewed increase in team interconnectivity and group cohesiveness, accompanied by a decrease in information propagation paths and centralized decision-making. Such a revival in collaborative efforts can be attributed to several factors, such as anniversary reflections, renewed public or scholarly interest, or ongoing impacts of the events. This pattern highlights the cyclical nature of collaboration in these domains, driven by evolving public and academic interest over time.

In summary, this relationship between artifact age and collaborative network dynamics sheds light on the adaptive nature of collaboration, particularly in areas marked by immediacy and long-term evolution. Understanding these patterns is crucial for managing and fostering effective collaborative environments over the lifespan of a project, especially in digital platforms where temporal dynamics play a significant role.

\textbf{\textit{The key finding regarding Hypothesis 2: team structures reveal evolving patterns over scale (size) and time (age). }}

However, the impact of these two factors—scale and time—on network structures manifests differently. In the technology domain, time did not show any significant impact on team structure, and in some domains such as entertainment, the scale did not show a significant impact on some of the team structure measures. Generally, the variation in network characteristics due to the aging of artifacts seems to be more uniform and less category-dependent as explained before. It should also be noted that the robustness of the findings of this paper is validated by testing lower time thresholds when calculating temporal weights of all the pairs of collaborators, using 24 hours instead of 48 hours, we observed consistent patterns in the data. This analysis underscores the stability of our results, even when subjected to more stringent temporal constraints. For additional details on this examination, readers are directed to the last section of the Appendix Document.

Our study's findings robustly validate Hypothesis 3, highlighting distinct collaborative network structures in digital platforms compared to traditional organizational settings. In this digital environment, we observed unique patterns of interconnectedness, clustering, and decentralization, contrasting with the more hierarchical structures prevalent in traditional organizations. The influence of artifact categories on network dynamics in digital contexts was particularly notable, underlining the importance of the nature of the artifact in shaping collaborative patterns. This contrast confirms the distinctiveness of digital collaboration, emphasizing the adaptability and specificity of digital platforms in facilitating collaborative processes.

\textbf{\textit{The key finding regarding Hypothesis 3: Team structure patterns in pure digital platforms drastically differ from traditional organizations. }}

\subsection{Network Dynamics and the category of the artifact}
In the realm of network analysis within organizational contexts, Jacobs and Watts have posited that "Network Heterogeneity Is Not Explained by Industry" \cite{jacobs2021large}. This perspective suggests that the categorization of firms by industry may be too broad a metric to significantly impact the network structures within them. Building upon this viewpoint, our research pivots to a more granular level of analysis, focusing on the nature of the artifact itself. This approach is driven by the assumption that the specific category of a topic might have a more direct influence on the structural characteristics of collaborative networks.

Our investigation, centered on the digital collaborative platform of Wikipedia, as detailed in the preceding sections, reveals a clear picture of how these network characteristics unfold across different topic domains. This exploration is vital, considering that the unique nature of each artifact category could potentially foster distinct collaborative patterns.

The results of our analysis present a fascinating narrative. The scaling of teams across different content categories reveals diverse and complex patterns in team structure. Conflict and disaster categories show a consistent trend towards more cohesive and structured collaboration with distributed leadership in larger teams. In contrast, politics and technology exhibit a more dynamic evolutionary process, with initial dispersal followed by increased cohesion and sub-community formation as teams grow. The entertainment domain, however, stands out with its more fluid and less predictable patterns, reflecting the adaptive nature of collaboration in this field. These insights offer a nuanced understanding of how team size influences collaboration, which is crucial for effectively managing and facilitating team dynamics in various contexts.

Similarly, when considering the age of artifacts, the conflicts and disasters categories again demonstrate a significant relationship with network characteristics. This could be attributed to the time-sensitive and evolving nature of these topics, where prolonged events necessitate and foster a deepening of collaborative networks over time. On the other hand, categories like politics and entertainment display a less consistent pattern, and the technology domain shows no meaningful pattern temporally, suggesting a multifaceted interplay of various factors influencing collaboration as these artifacts mature. 

\textbf{\textit{The key finding regarding Hypothesis 1: Problem context matters in the team structure.}}

\section{Conclusion}

In an era marked by rapid technological advancement and globalization, social dynamics, facilitated by technology, adapt and evolve within sociotechnical systems. This adaptation is particularly evident in digital collaborative environments, where technology not only supports but also transforms how knowledge is created and shared. In our study, we implemented a novel approach to network construction, tailored to capture the intricate dynamics of collaboration within digital platforms. This methodology stands out due to its innovative integration of both temporal and content-based interactions among users. By systematically analyzing edit histories across various time windows and evaluating user interactions not just in terms of alternating interactions restricted by a time threshold (ATI) but also content overlap, our approach offers a more multidimensional perspective on collaborative behaviors. This method allows for a deeper understanding of how digital collaborations evolve over time and how content-related factors influence the formation and dynamics of collaborative networks. Our study, on the backdrop of an increasingly digitalized organizational landscape, reveals critical insights into the dynamics of digital collaboration, pertinent to an era where teamwork transcends traditional communication methods.

The findings corroborate three central hypotheses, each shedding light on distinct aspects of digital team behaviors: firstly, validating Hypothesis 1, we found that the context of a problem significantly influences team structures on digital platforms. This underlines the importance of considering the specific nature of tasks or topics when analyzing collaborative dynamics, which is a critical factor in shaping how teams form and function in digital environments.  In addressing Hypothesis 2, our study unveils that team structures on digital platforms reveal evolving patterns over time and scale. The insights gained here are instrumental for understanding how digital teams grow and evolve, providing key considerations for managing and optimizing these evolving networks. Hypothesis 3 finds strong validation in our analysis, indicating that team structure patterns in purely digital platforms differ markedly from those in traditional organizational settings. This finding is crucial in an era where digital platforms are increasingly harnessed for diverse purposes, from knowledge sharing to collective product and service design. Understanding these unique digital collaboration dynamics is imperative for high-level leaders and decision-makers of such platforms.

Our study underscores the importance of decentralized leadership in digital teams, offering insights to enhance knowledge sharing. By mapping out collaborative patterns, particularly in platforms like Wikipedia, this research paves the way for strategic interventions in digital teams. Such interventions are crucial for aligning team dynamics with broader organizational goals, thereby optimizing the efficacy of digital collaborations. In conclusion, this research provides a comprehensive snapshot of digital collaboration dynamics, serving as a foundational step toward a deeper understanding and optimization of digital team interactions. As the landscape of team collaboration continues to evolve, the insights from this study will be invaluable for navigating the complexities of teamwork in the digital era, enabling organizations to harness the full potential of their digital collaborative endeavors.

\section{Limitations and Future Directions}

Although our study offers significant insights into the collaborative dynamics within Wikipedia, it's important to acknowledge its limitations and propose research paths that could further deepen our understanding of digital collaboration. This study is focused on a single open-source digital platform suggesting the need for broader explorations. A promising direction for future research is “a cross-platform comparative analysis”, which would involve examining and contrasting collaborative patterns across different platforms. This could determine if the distinctive collaborative behaviors identified on Wikipedia hold true in other digital contexts, characterized by their own interfaces, user bases, and objectives, providing a broader perspective across various digital environments.

Another crucial path for future research lies in the “behavioral aspects of digital collaboration”. In the digital collaboration landscape, factors like trust and collective psychological ownership are paramount. As \cite{sensuse2021exploring} emphasizes, trust is foundational to effective collaboration and knowledge sharing. Positive team interactions have been shown to significantly enhance trust levels. Similarly, the concept of collective psychological ownership, as introduced by Gray and colleagues, is increasingly recognized as integral to the success and creativity of teams \cite{gray2020emergence}, especially in digital settings where a sense of shared responsibility and investment is crucial. These factors, pivotal in the realm of digital collaboration, are both supported by digital platforms' feature of traceability. Every step of the knowledge creation and collaboration process is documented; however, there remains a gap in fully understanding how these factors play out in digital contexts. Future research should focus on how digital platforms can optimize elements like trust and collective ownership and, in turn, how these elements influence team effectiveness and innovation capacity.

In conclusion, as the landscape of collaboration evolves, with an increasing shift towards purely online methods juxtaposed against traditional models, the need for continuous experimentation, analysis, and research in this area becomes ever more critical. This study underscores the evolving landscape of digital collaboration, revealing critical insights into the impact of problem context, team size, and problem maturity on collaborative network structures. Demonstrating significant differences between digital platforms and traditional organizational models, this study’s insights are pivotal for organizations embracing digital platforms, guiding them towards more effective team structures and enhanced creativity as sociotechnical systems. The study not only enriches the academic discourse on digital collaboration but also offers practical applications for organizations striving to harness the full potential of digital collaborative environments in achieving their strategic goals and fostering innovation in an increasingly online world.

{\appendix[Complementary Document]
[The link to the Appendix Document will be added here.]}


%

\printbibliography

\begin{IEEEbiography}[{\includegraphics[width=1in,height=1.25in,clip,keepaspectratio]{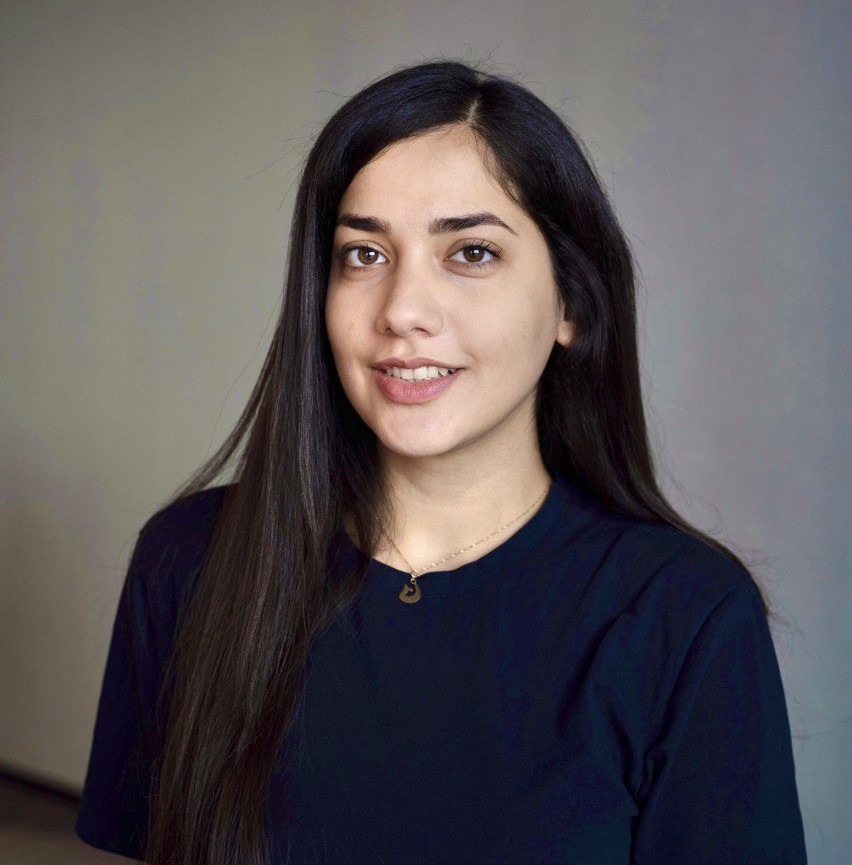}}]{Negin Maddah}
received a B.S. degree in Industrial Engineering from Alzahra University, Tehran, Iran in 2017, and an M.S. in Industrial Engineering, Systems Optimizations from K. N. Toosi University of Technology, Tehran, Iran in 2020. She is currently a Ph.D candidate, pursuing a Ph.D. degree in Industrial Engineering in the Multi-AGent Intelligent Complex Systems (MAGICS) Lab at Northeastern University, Boston, MA. Her research interests include complex systems, computational social sciences, sociotechnical systems, platform-based systems, and data science.
\end{IEEEbiography}

\begin{IEEEbiography}[{\includegraphics[width=1in,height=1.25in,clip,keepaspectratio]{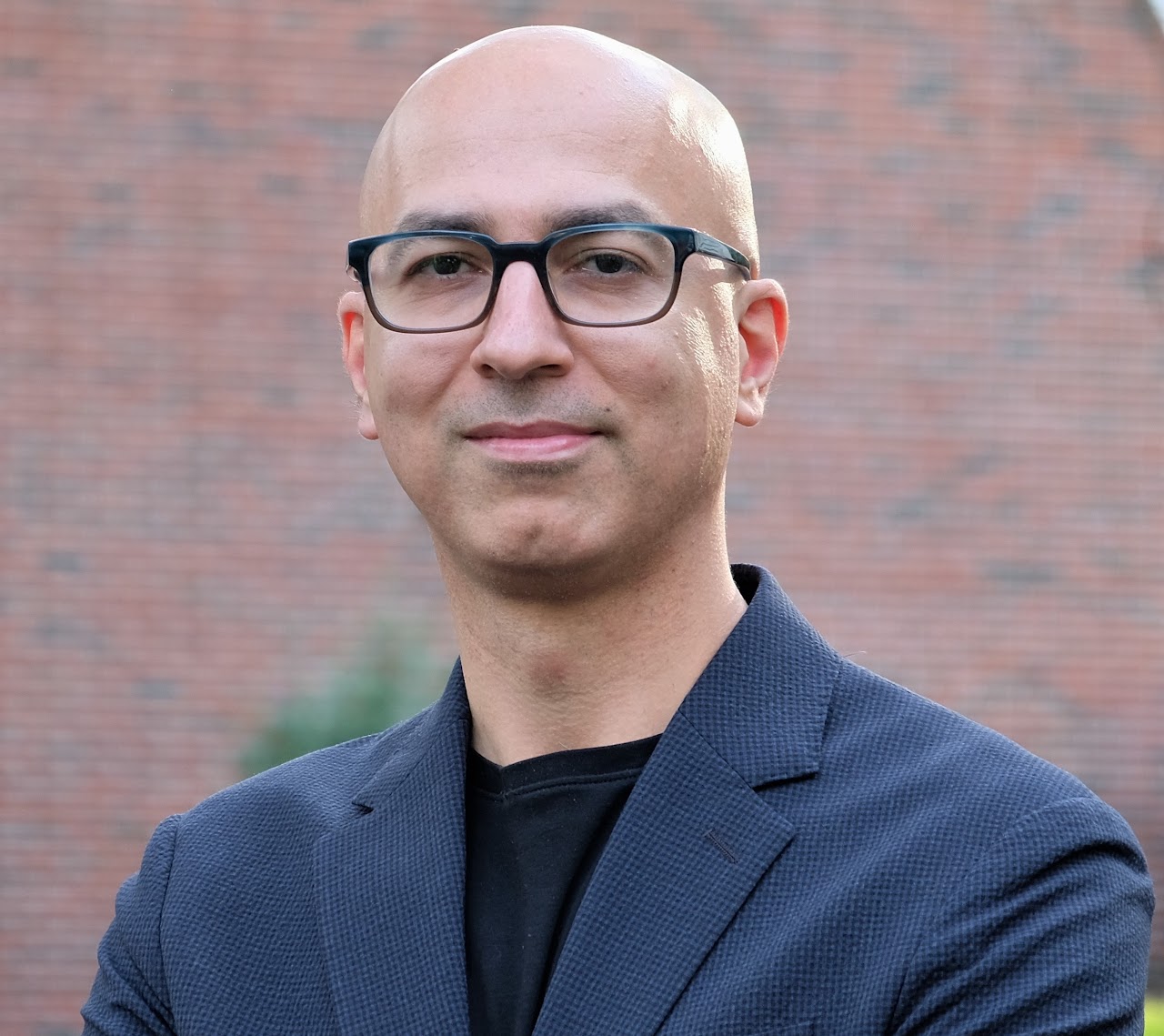}}]{Babak Heydari} earned his PhD from the University of California at Berkeley. He is an Associate Professor in the Department of Mechanical and Industrial Engineering and Co-Program Director of the Master of Science in Engineering Management program. Babak Heydari is also the Director of the Multi-AGent Intelligent Complex Systems (MAGICS) Lab and affiliate faculty at the School of Public Policy and Urban Affairs and the Network Science Institute. His research interests include sociotechnical systems, artificial intelligence, systems engineering and design, social and economic networks, resilience of complex systems, computational social sciences, platform-based systems, sharing economy, and computational social sciences.
\end{IEEEbiography}

\vspace{11pt}

\vfill

\end{document}

%% file: table1.tex
\begin{table}
\caption{Summary of Constructed Networks by Category}
\label{tab:table1}
\resizebox{\columnwidth}{!}{%
\begin{tabular}{@{}ccccccc@{}}
\toprule
Category             & \#Topics             & \#Networks           & Min Start Date                                                       & Max End Date                                                         & \begin{tabular}[c]{@{}c@{}}Min \\ \#Nodes\end{tabular} & \begin{tabular}[c]{@{}c@{}}Max \\ \#Nodes\end{tabular} \\ \midrule
Politics             & 12                   & 148                  & \begin{tabular}[c]{@{}c@{}}2011-11-28 \\ 00:18:34+00:00\end{tabular} & \begin{tabular}[c]{@{}c@{}}2023-11-30 \\ 22:57:44+00:00\end{tabular} & 2                                                      & 29                                                     \\
\multicolumn{1}{l}{} & \multicolumn{1}{l}{} & \multicolumn{1}{l}{} & \multicolumn{1}{l}{}                                                 & \multicolumn{1}{l}{}                                                 & \multicolumn{1}{l}{}                                   & \multicolumn{1}{l}{}                                   \\
Conflicts            & 12                   & 155                  & \begin{tabular}[c]{@{}c@{}}2014-03-01 \\ 23:01:38+00:00\end{tabular} & \begin{tabular}[c]{@{}c@{}}2023-12-23 \\ 17:07:20+00:00\end{tabular} & 2                                                      & 27                                                     \\
\multicolumn{1}{l}{} & \multicolumn{1}{l}{} & \multicolumn{1}{l}{} & \multicolumn{1}{l}{}                                                 & \multicolumn{1}{l}{}                                                 & \multicolumn{1}{l}{}                                   & \multicolumn{1}{l}{}                                   \\
Disasters            & 14                   & 144                  & \begin{tabular}[c]{@{}c@{}}2009-10-25 \\ 21:51:24+00:00\end{tabular} & \begin{tabular}[c]{@{}c@{}}2023-12-10 \\ 21:02:50+00:00\end{tabular} & 2                                                      & 27                                                     \\
\multicolumn{1}{l}{} & \multicolumn{1}{l}{} & \multicolumn{1}{l}{} & \multicolumn{1}{l}{}                                                 & \multicolumn{1}{l}{}                                                 & \multicolumn{1}{l}{}                                   & \multicolumn{1}{l}{}                                   \\
Tech                 & 12                   & 176                  & \begin{tabular}[c]{@{}c@{}}2010-01-03 \\ 08:46:49+00:00\end{tabular} & \begin{tabular}[c]{@{}c@{}}2023-11-02 \\ 21:03:38+00:00\end{tabular} & 2                                                      & 31                                                     \\
\multicolumn{1}{l}{} & \multicolumn{1}{l}{} & \multicolumn{1}{l}{} & \multicolumn{1}{l}{}                                                 & \multicolumn{1}{l}{}                                                 & \multicolumn{1}{l}{}                                   & \multicolumn{1}{l}{}                                   \\
Entertainment        & 13                   & 166                  & \begin{tabular}[c]{@{}c@{}}2010-06-03 \\ 00:19:30+00:00\end{tabular} & \begin{tabular}[c]{@{}c@{}}2023-12-18 \\ 13:54:27+00:00\end{tabular} & 2                                                      & 36                                                     \\ \bottomrule
\end{tabular}%
}
\end{table}

%% file: tablesize.tex
\begin{table}
\caption{Regression Analysis of Network Structure Metrics and Team Size}
\label{tab:size-table}
\resizebox{\columnwidth}{!}{%
\begin{tabular}{@{}cccccc@{}}
\toprule
Model Variables                                                                                 & Category      & \begin{tabular}[c]{@{}c@{}}Size \\ p-value\end{tabular} & \begin{tabular}[c]{@{}c@{}}\(Size^2\)\\ P-value\end{tabular} & \begin{tabular}[c]{@{}c@{}}Prob\\ (F-statistics)\end{tabular} & \begin{tabular}[c]{@{}c@{}}Adjusted\\ \(R^2\)\end{tabular} \\ \midrule
\multirow{5}{*}{\begin{tabular}[c]{@{}c@{}}Average Degree\\  vs Team Size\end{tabular}}         & Politics      & 0.672                                                   & $\ll 0.001*$                                               & $\ll 0.001*$                                  & 0.548                                                 \\
                                                                                                & Conflicts     & $\ll 0.001*$                             & 0.768                                                                     & $\ll 0.001*$                                   & 0.480                                                 \\
                                                                                                & Disasters     & 0.016*                                                  & 0.430                                                                     & $\ll 0.001*$                                   & 0.483                                                 \\
                                                                                                & Tech          & $\ll 0.001*$                             & 0.005*                                                                    & $\ll 0.001*$                                   & 0.490                                                 \\
                                                                                                & Entertainment & 0.027                                                   & $\ll 0.001*$                                               & $\ll 0.001*$                                  & 0.571                                                 \\
\multirow{5}{*}{\begin{tabular}[c]{@{}c@{}}Clustering\\ vs Team Size\end{tabular}}              & Politics      & $\ll 0.001*$                            & 0.005*                                                                    & $\ll 0.001*$                                  & 0.191                                                 \\
                                                                                                & Conflicts     & $\ll 0.001*$                             & 0.001*                                                                    & $\ll 0.001*$                                   & 0.163                                                 \\
                                                                                                & Disasters     & $\ll 0.001*$                             & 0.002*                                                                    & $\ll 0.001*$                                   & 0.187                                                 \\
                                                                                                & Tech          & $\ll 0.001*$                             & 0.016*                                                                    & $\ll 0.001*$                                   & 0.084                                                 \\
                                                                                                & Entertainment & 0.874                                                   & 0.293                                                                   &  $\ll 0.001*$                                   & 0.109                                                 \\
\multirow{5}{*}{\begin{tabular}[c]{@{}c@{}}Average Shortest\\ Path\\ Vs Team Size\end{tabular}} & Politics      & $\ll 0.001*$                            & $\ll 0.001*$                                              & $\ll 0.001*$                                   & 0.382                                                 \\
                                                                                                & Conflicts     & $\ll 0.001*$                             & $\ll 0.001*$                                               & $\ll 0.001*$                                   & 0.389                                                 \\
                                                                                                & Disasters     & $\ll 0.001*$                             & $\ll 0.001*$                                               & $\ll 0.001*$                                   & 0.511                                                 \\
                                                                                                & Tech          & $\ll 0.001*$                             & 0.003*                                                                    & $\ll 0.001*$                                   & 0.600                                                 \\
                                                                                                & Entertainment & $\ll 0.001*$                             & $\ll 0.001*$                                              & $\ll 0.001*$                                   & 0.217                                                 \\
\multirow{5}{*}{\begin{tabular}[c]{@{}c@{}}Centralization\\ Vs Team Size\end{tabular}}          & Politics      & 0.349                                                   & 0.047*                                                                    & $\ll 0.001*$                                   & 0.086                                                 \\
                                                                                                & Conflicts     & 0.303                                                   & 0.069                                                                     & 0.011*                                                        & 0.045                                                 \\
                                                                                                & Disasters     & 0.018*                                                  & 0.008*                                                                    & 0.023*                                                        & 0.040                                                 \\
                                                                                                & Tech          & 0.427                                                   & 0.178                                                                     & 0.120                                                         & 0.013                                                 \\
                                                                                                & Entertainment & 0.015*                                                  & 0.561                                                                   &  $\ll 0.001*$                                  & 0.244                                                 \\ \cmidrule{1-6} 
\end{tabular}%
}
\end{table}

%% file: tableage.tex
\begin{table}
\caption{Regression Analysis of Network Structure Metrics and Artifact Age}
\label{tab:age-table}
\resizebox{\columnwidth}{!}{%
\begin{tabular}{@{}cccccc@{}}
\toprule
Model Variables                                                                                    & Category      & \begin{tabular}[c]{@{}c@{}}Age\\ p-value\end{tabular} & \begin{tabular}[c]{@{}c@{}}\(Age^2\)\\ P-value\end{tabular} & \begin{tabular}[c]{@{}c@{}}Prob\\ (F-statistics)\end{tabular} & \begin{tabular}[c]{@{}c@{}}Adjusted\\ \(R^2\)\end{tabular} \\ \midrule
\multirow{5}{*}{\begin{tabular}[c]{@{}c@{}}Average Degree\\  vs Artifact Age\end{tabular}}         & Politics      & 0.003*                                                & 0.005*                                                                   & 0.013*                                                         & 0.051                                                 \\
                                                                                                   & Conflicts     & $\ll 0.001*$                           & $\ll 0.001*$                                              & $\ll 0.001*$                                  & 0.280                                                 \\
                                                                                                   & Disasters     & $\ll 0.001*$                          & 0.003*                                                                   & $\ll 0.001*$                                   & 0.124                                                 \\
                                                                                                   & Tech          & 0.079                                                 & 0.140                                                                    & 0.157                                                         & 0.011                                                \\
                                                                                                   & Entertainment & $\ll 0.001*$                           & $\ll 0.001*$                                              & $\ll 0.001*$                                   & 0.081                                                 \\
\multirow{5}{*}{\begin{tabular}[c]{@{}c@{}}Clustering\\ vs Artifact Age\end{tabular}}              & Politics      & 0.039*                                                & 0.124                                                                    & 0.050*                                                        & 0.029                                                 \\
                                                                                                   & Conflicts     & $\ll 0.001*$                           & 0.002*                                                                   & 0.000*                                                        & 0.111                                                 \\
                                                                                                   & Disasters     & 0.005*                                                & 0.004*                                                                   & 0.014*                                                        & 0.075                                                 \\
                                                                                                   & Tech          & 0.159                                                 & 0.098                                                                    & 0.220                                                         & 0.008                                                 \\
                                                                                                   & Entertainment & 0.007*                                                & 0.021*                                                                   & 0.019*                                                        & 0.037                                                 \\
\multirow{5}{*}{\begin{tabular}[c]{@{}c@{}}Average Shortest\\ Path\\ Vs Artifact Age\end{tabular}} & Politics      & 0.847                                                 & 0.918                                                                    & 0.687                                                         & 0.000                                                 \\
                                                                                                   & Conflicts     & 0.017*                                                & 0.017*                                                                   & 0.050*                                                        & 0.031                                                 \\
                                                                                                   & Disasters     & 0.004*                           & 0.007*                                                                   & 0.016*                                                        & 0.072                                                 \\
                                                                                                   & Tech          & 0.221                                                 & 0.208                                                                    & 0.450                                                         & 0.000                                                 \\
                                                                                                   & Entertainment & 0.248                                                 & 0.176                                                                    & 0.376                                                         & 0.000                                                \\
\multirow{5}{*}{\begin{tabular}[c]{@{}c@{}}Centralization\\ Vs Artifact Age\end{tabular}}          & Politics      & 0.269                                                 & 0.402                                                                    & 0.451                                                         & -0.003                                                \\
                                                                                                   & Conflicts     & $\ll 0.001*$                           & 0.002*                                                                   & $\ll 0.001*$                                   & 0.115                                                 \\
                                                                                                   & Disasters     & 0.007*                                                & 0.009*                                                                   & 0.026*                                                        & 0.061                                                 \\
                                                                                                   & Tech          & 0.100                                                 & 0.128                                                                    & 0.254                                                         & 0.006                                                \\
                                                                                                   & Entertainment & 0.021*                                                & 0.077                                                                    & 0.034*                                                        & 0.029                                                 \\ \cmidrule{1-6} 
\end{tabular}%
}
\end{table}